# Fiber-pigtailing quantum-dot cavity-enhanced light emitting diodes

L. Rickert,[1] F. Schröder,[1] T. Gao,[1] C. Schneider,[2,3] S. Höfling,[2] and T. Heindel[1,a]

[1]*Institut für Festkörperphysik, Technische Universität Berlin, 10623 Berlin, Germany*

[2]*Technische Physik, Physikalisches Institut and Wilhelm Conrad Röntgen Research Center for Complex Material Systems, Universität Würzburg, 97074 Würzburg, Germany*

[3]*Institut für Physik, Carl von Ossietzky Universität Oldenburg, 26129 Oldenburg, Germany*

a) Author to whom correspondence should be addressed: tobias.heindel@tu-berlin.de

We report on a process for the fiber-coupling of electrically driven cavity-enhanced quantum dot light emitting devices. The developed technique allows for the direct and permanent coupling of p-i-n-doped quantum dot micropillar cavities to single-mode optical fibers. The coupling process, fully carried out at room temperature, involves a spatial scanning technique, where the fiber facet is positioned relative to a device with a diameter of 2 µm using the fiber-coupled electroluminescence of the cavity emission as a feedback parameter. Subsequent gluing and UV curing enables a rigid and permanent coupling between micropillar and fiber core. Comparing our experimental results with finite element method simulations indicates a cavity-to-fiber mode-coupling efficiency of ~46%. Furthermore, we demonstrate pulsed current injection at a repetition rate exceeding 200 MHz as well as low-temperature operation down to 77 K of the fiber-coupled micropillar device. The technique presented in this work is an important step in the quest for efficient and practical quantum light sources for applications in quantum information.

Solid-state based quantum-light sources are elementary building blocks for photonic quantum technologies [1-3]. Specifically, the maturity of single-photon sources (SPSs) based on semiconductor quantum dots (QDs) has advanced substantially in recent years [4,5], allowing for the efficient generation of quantum states of light under optical [6-9] or electrical [10,11] pumping. As a result, QD-based quantum light sources have been employed for many proof-of-concept experiments on quantum communication [12-14] and photonic quantum computing [15]. Most of these experiments, however, suffer from rather complex and bulky setups due to complex light extraction via free-space optics, hindering more widespread applications. On the other hand, the development of user-friendly devices for applications outside shielded lab environments recently attracted much interest [16,17]. A crucial aspect in this context concerns the direct coupling of the quantum light sources to optical single-mode (SM) fibers facilitating a robust packaging of the devices. Pioneering work in this direction utilized fiber-coupled QD samples employing fiber-bundles containing about 600 individual fiber cores to spatially post-select a single emitter [18]. More recently, the direct fiber-coupling of optically pumped photonic nanostructures with embedded QDs, such as photonic wires [19] and micropillars with oxide aperture [20], has been reported. The latter scheme has also been used to realize an optically pumped cavity-enhanced single-photon source with gates for spectral tuning of the QD emission [21]. In addition, optically-pumped fiber-integrated microcavities were employed for the generation of coherent acoustic phonons [22]. Moreover, schemes for the SM fiber-coupling of QD microlenses are currently under development using two-photon direct laser writing of photonic microstructures [23]. While these previous efforts relied entirely on optically excited QDs, an electrically driven directly fiber-coupled QD device, not demonstrated to date, would clearly represent an important step in device integration toward efficient and practical quantum light sources.

In this work, we developed a technique enabling the direct and permanent coupling of electrically-pumped cavity-based quantum light sources to optical SM fibers. Our process exploits the emission of the cavity modes as direct feedback parameter for the precise relative alignment between cavity and fiber core. As all steps of this process can be carried out at room temperature, a simple setup can be used for its implementation.

The fiber-pigtailed device fabricated in this work is illustrated in Fig. 1(a). It comprises an electrically contacted QD micropillar cavity aligned with the core of a SM fiber (Thorlabs 780HP, core diameter $d_{core}$ = 4.4 µm) mounted in a ceramic ferrule (not shown in Fig. 1(a)) serving as larger contact area for the gluing step described further below. The micropillar cavity device is based on a cylindrical Fabry-Pérot resonator which is composed of an intrinsic λ-thick GaAs-cavity with a layer of self-organized InAs QDs in its center and a lower n-doped and an upper p-doped distributed Bragg reflector (DBR). The number of the AlAs/GaAs mirror pairs is optimized for high photon extraction efficiencies resulting in moderate Q factors (~1000-3000) for devices of $d_{pillar}$ = 2.0 µm diameter. The micropillar is planarized using benzocyclobutene to support the ring-shaped electrical top contact, which is used for peripheral current injection together with a planar back contact. For details on the device design and fabrication we refer the interested reader to Ref. [24] and to Ref. [25] for the demonstration of an efficient electrically triggered single-photon source. An important characteristic of the sample used for the experiments in the following, refers to their spectral properties at low- and room-temperature, respectively. The QD-micropillars are deliberately designed such, that at low temperature (~ 4 K) the emission of the fundamental cavity mode is spectrally detuned (by about -25 nm) with respect to the QD-ensemble (full-width at half-maximum ~ 25 nm). This results in a low spectral density of QD-emission lines close to the fundamental cavity mode, which enabled the observation of pronounced cavity-enhanced single-photon emission in previous reports (cf. Ref. [25]). At room temperature, both the emission of the cavity-modes and the QD ensemble are shifted to longer wavelengths. Due to the stronger temperature-dependence for the QD's emission energy, however, the spectral detuning between QD-ensemble and cavity-mode is reduced compared to 4 K. The resulting higher spectral density of the QD ensemble in the vicinity of the fundamental cavity mode can be exploited for the fiber-coupling process presented in the following. Noteworthy, we choose a small fiber ferrule in our process with a diameter of 1.25 mm (standard size: 2.5 mm) for two reasons: First, the chip containing the target device can be made smaller, resulting in a larger yield of pigtailed devices per wafer. Second, the distance between target device and wire bond used for current injection can be shorter and damage of the wire bond during alignment of the ferrule is less probable.

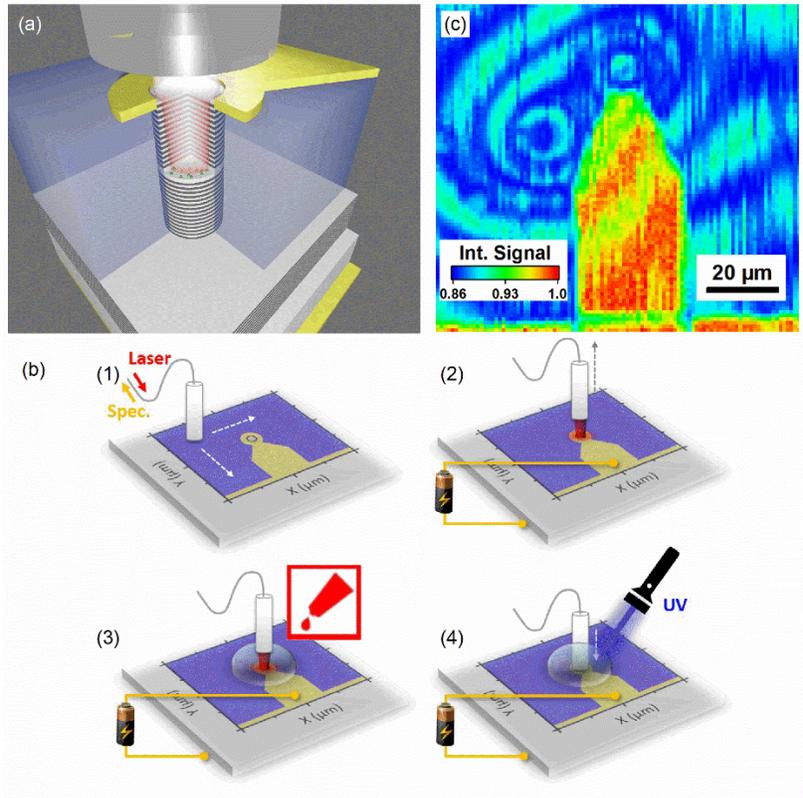

Figure 1. (a) Schematic representation of the fiber-pigtailed electrically contacted QD-micropillar. (b) Illustration of the fiber-coupling process conducted at room temperature: (1) scanning of the fiber-coupled laser signal to locate the target pillar, (2) precise fiber-to-pillar alignment via direct monitoring of the cavity's electroluminescence signal, (3) application of UV adhesive, and (4) UV curing using the cavity electroluminescence as feedback parameter. (c) Integrated intensity map of the reflected laser emission (spectral range: 655-660 nm) of a $100 \times 100$ µm$^2$ scan with 1 µm step size during step (1) of the fiber-coupling process.

For the precise coupling of QD-micropillar and SM fiber, we developed a robust four-step process fully performed under ambient conditions (i.e., room temperature, no vacuum) as sketched in Fig. 1(b). In step (1), the initial alignment of the fiber ferrule relative to the electrical contacts takes place using a laser-scanning method. Here, we detect the fiber-coupled emission of a diode laser (λ = 657 nm) after backreflection at the sample surface through a 90:10 fiber beam splitter using a spectrometer (10% excitation path, 90% detection path), while scanning the fiber facet across the sample surface using a 3D closed-loop piezo-stage. The resulting map of the integrated reflected intensity, exemplarily depicted in Fig. 1(c) for a micropillar with $d_{pillar}$ = 4.0 µm, enables one to locate the top contacts as well as the top facet of a target micropillar. In step (2) the fiber core is precisely aligned with the micropillar to maximize the mode coupling by using the fiber-coupled cavity emission (~900 nm) of the micropillar driven under direct current (dc) injection as a feedback parameter. At this point, the distinct spectral characteristics of the QD-micropillar sample comes into play (cf. sample description), which enable the illumination of the cavity-modes via the emission of the QD-ensemble at room temperature. Compared to alignment under purely optical excitation, this procedure has the additional advantage that the integrity of the electrical contacts can be monitored throughout the alignment process as well as the subsequent gluing and curing step. The height of the scanning fiber just before contact is adjusted by tracking the spectral position of the fundamental cavity mode, which shows a spectral blue-shift once the ferrule starts to strain the micropillar. Additionally, the QD-micropillar's far-field emission properties can be studied during this step, as discussed further below (cf. Fig. 2). Step (3) concerns the application of optical UV-sensitive adhesive (NOA81, $n = 1.56$): After maximizing the coupling of the cavity emission to the fiber mode, the fiber is lifted by about 30 mm normal to the sample surface, for applying the optical adhesive onto the sample using a syringe. Subsequently, the fiber is lowered again to its previous position. After optimizing the coupling once again, the UV curing process is started in step (4). Several exposure steps with the increasing dose of collimated UV radiation (spot size 9 mm) between 10 and 160 mW/cm² were applied over several minutes to achieve complete curation of the adhesive. As mentioned earlier, the cavity emission is monitored during this process. After UV curing, the device is additionally encapsulated using a two-component epoxy glue for further mechanical stability. The EL emission properties of the QD-micropillar during and after the UV curing of the coupling process are discussed further below (cf. Fig. 3).

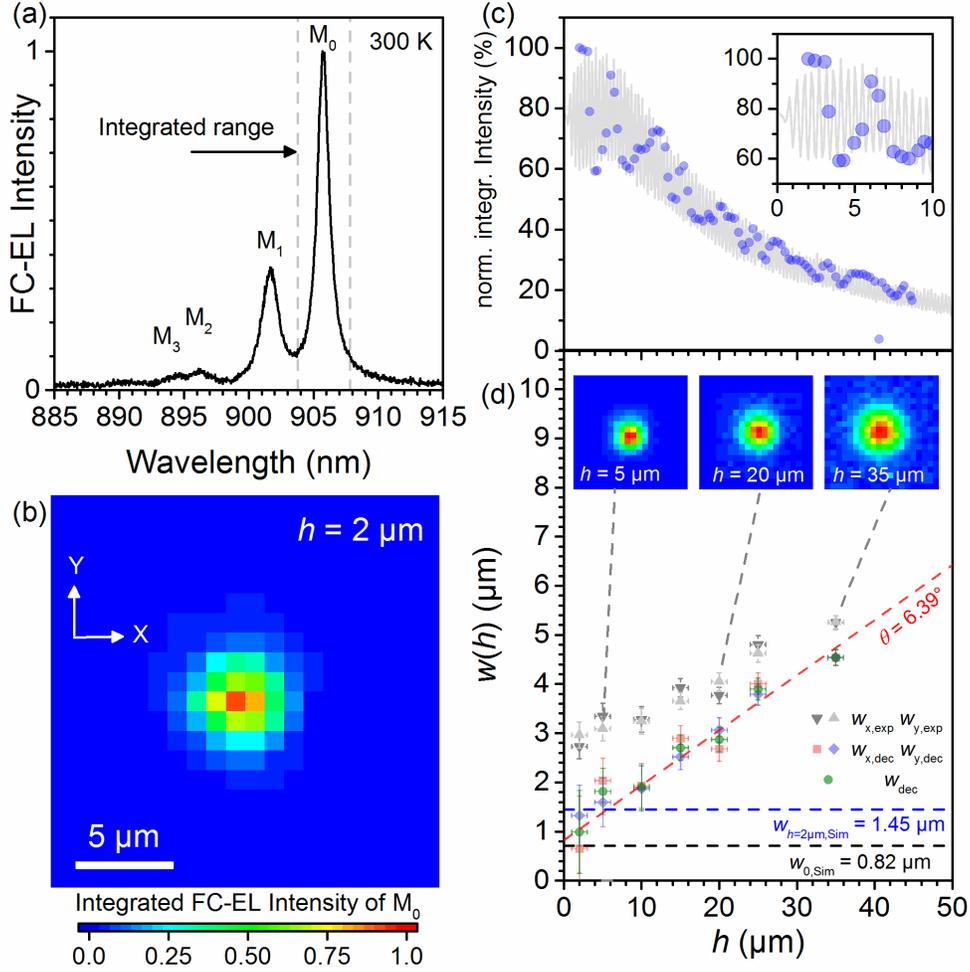

Figure 2. Investigation on the far-field emission of a QD micropillar ($d_{pillar}$ = 2 µm) directly coupled to a SM fiber before gluing. (a) Exemplary fiber-coupled electroluminescence (FC-EL) spectrum of the QD-micropillar operated under dc injection ($I_{dc}$ = 16 µA) at room temperature. (b) 20 × 20 µm² far-field map of the integrated $M_0$ mode emission at 905.8 nm collected through the SM fiber. (c) Measured integrated $M_0$ mode FC-EL intensity (blue dots) and simulated fiber mode coupling efficiency (gray line) as function of the fiber-to-pillar distance $h$. Both simulated and experimental data are normalized to their respective maximum values to allow direct comparison. (d) Mode field radius $w$ of the far-field emission of $M_0$ as a function of $h$. Symbols refer to the far-field emission radius in x- and y-direction as measured (light/dark gray triangles) and deconvoluted with the fiber mode-field radius (red/blue squares). The horizontal dashed lines indicate mode radii obtained from FEM simulations at fiber-pillar distances $h$ = 0 and 2 µm. The divergence angle $\theta$ is extracted from the slope of the linear regression. Exemplary far-field emission patterns underlying the experimental data in (d) are shown for $h$ = 5, 20, and 35 µm.

During the fiber-coupling process described above, the sample is mounted on a custom-made chip carrier with pin-connectors, bond pads for wire bonding, and a copper heatsink assuring thermalization of the micropillar sample for future operation at cryogenic temperatures. A coarse positioning of the micropillar device relative to the fiber is performed with an x-y-stage with a resolution of 1 µm, while a 3D piezo-stage is used for fine positioning of the fiber ferrule with a resolution of better than 250 nm. Moreover, the fiber-to-micropillar distance can be coarsely adjusted with a z-stage (resolution: 1 µm). For electrical operation of the QD micropillar chip, we use a source measurement unit or a pulse pattern generator (300 ps minimal pulse-width, 250 MHz maximal repetition rate) for dc or pulsed current injection, respectively. The emission of the fiber-pigtailed microcavity is spectrally analyzed using a spectrometer with a spectral resolution of 25 µeV.

For the final device, we choose a QD micropillar with a $d_{\text{pillar}} = 2.0$ µm, being the optimal trade-off between spectral QD density and photon extraction efficiency for this type of sample. Figure 2(a) shows a spectrum of the SM fiber coupled emission of the target device operated at room temperature and under dc injection ($I = 16$ µA, 5 s integration time). The spectrum was recorded during the alignment step (2) with the fiber facet being close to surface contact. Emission of the fundamental cavity mode ($M_0$) at a wavelength of 905.8 nm as well as several higher order modes $M_{1-3}$ are clearly visible. The far-field pattern of the $M_0$-mode is spatially resolved in the map shown in Fig. 2(b) displaying the fiber-coupled electroluminescense (FC-EL) of the integrated mode intensity. Figure 2(c) depicts the integrated FC-EL intensity of mode $M_0$ extracted from the corresponding far-field patterns as a function of $h$. Here, we calibrated the origin of the $h$-axis to the point, at which the emission of a neighboring device started to show a blue-shift in the mode-spectra, indicating the point of physical contact.

The integrated intensity decreases with increasing $h$ due to the finite beam-divergence of the micropillar cavity, which will be investigated in detail further below. Pronounced oscillations are observed, which are due to the interference between the facets of micropillar and fiber core. This observation is in excellent quantitative agreement with the calculated mode coupling efficiency (gray line) obtained from finite element methods (FEM) simulations using the software package JCMsuite [26], assuming a dipole source at the center of the micropillar cavity. Note that we are not able to resolve the period of the interference fringes in our experiment, due to the limited resolution of the z-stage used (1 µm). To investigate the far-field emission properties of the micropillar cavity under study in more detail, we recorded FC-EL maps under fiber-coupled conditions as a function of $h$. From the recorded far-field patterns, we extracted the mode-field radius of the fundamental cavity mode $M_0$ by fitting an elliptical 2D Gaussian function to the far-field data. For the data recorded at $h = 2$ µm, we extract mode-field radii of $w_{x,exp} = (2.73 \pm 0.25)$ µm and $w_{y,exp} = (2.96 \pm 0.27)$ µm in x- and y-direction, respectively. Considering the finite mode-field radius of the fiber of $w_{fiber} = 2.65$ µm used to probe the far-field emission pattern, we obtain deconvoluted values according to $w_{dec} = \sqrt{w_{exp}^2 - w_{fiber}^2}$ of $w_{x,dec.} = (0.66^{+1.06}_{-0.66})$ µm and $w_{y,dec} = (1.32 \pm 0.61)$ µm. Here, the large errors cannot be avoided, due to the strong convolution of the mode-field radii of fiber and micropillar cavity. The deconvoluted mode radii as function of $h$ are displayed in Fig. 2(d). Additionally, far-field EL emission intensity maps are displayed for $h = 5, 20$ and $35$ µm.

The averaged value of $w_{dec} = (w_{x,dec} + w_{y,dec})/2 = (0.99 \pm 0.84)$ µm at $h = 2$ µm lies well within the range expected from the FEM simulations, revealing a beam radius between $w_{h=2\mu m,sim} = 1.45$ µm and $w_{0,sim} = 0.82$ µm at distances of at $h = 2$ µm and 0 µm, respectively. The beam-divergence of the fundamental mode is directly observed in Fig. 2(d), and we obtain a divergence angle $\theta = (6.4 \pm 1.0)°$ from the mode field radii ($(3.7 \pm 0.7)°$ from radius of half maximum). These values are somewhat smaller compared to the divergence angles reported in previous experimental reports on optically pumped micropillars of comparable size [27], a discrepancy that we explain by the differently structured devices (e.g. electrical ring-contact and different heights of the micropillar). From the values for the mode-field radii determined above, we can estimate the mode coupling efficiency or power transmission coefficient of the fundamental cavity-mode emission to the SM fiber in our experiment as $\eta = \left(\frac{2w_{pillar}w_{fiber}}{w_{pillar}^2 + w_{fiber}^2}\right)^2 \exp\left(-\frac{2u^2}{w_{pillar}^2 + w_{fiber}^2}\right)$, where $u$ denotes the transverse misalignment accounting for our positioning accuracy [21,28]. We choose this quantity η to quantify the coupling efficiency, as it is directly accessible during our fiber-pigtailing process. In future work, also the overall device efficiency is an important measure to determine. Assuming a deconvoluted pillar mode radius obtained from a radial symmetric 2D-Gaussian of $w_{pillar} = (1.05 \pm 0.51)$ µm, we obtain an estimated mode coupling efficiency of $\eta = (46 \pm 33)$%, which lies again within the theoretically expected range between $(31.45 \pm 0.83)$% and $(69.96 \pm 0.80)$% at $h = 0$ µm and $h = 2$ µm, respectively, obtained with the simulated beam radii $w_{0,Sim}$ and $w_{h=2\mu m,Sim}$ described above. Again, the large error in the estimation of $\eta$ based on experimental parameters is caused by the relatively large fiber mode-field diameter used to probe the micropillar emission with a significantly smaller mode-field diameter.

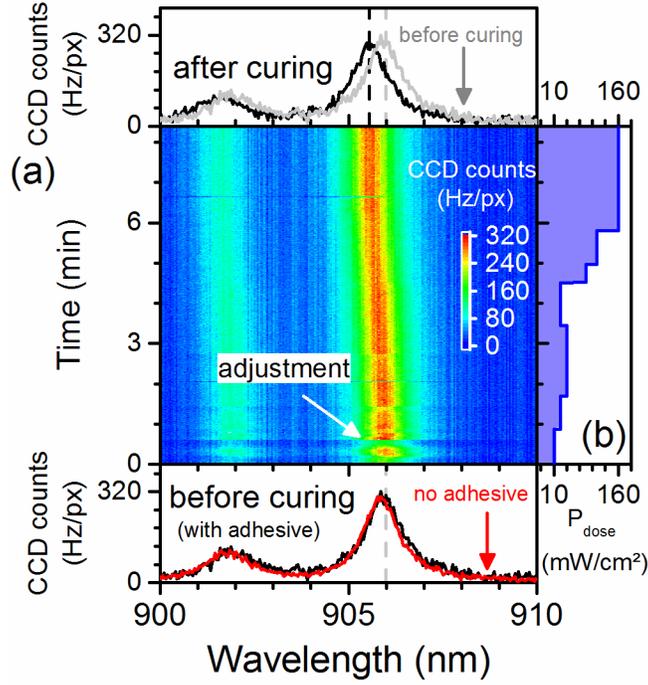

Figure 3. (a) FC-EL emission spectrum of the target micropillar at the beginning (with and without applied UV adhesive), during, and after the UV curing process. (b) Applied UV illumination dose $P_{\text{dose}}$ during the curing process.

Next, we proceed with step (4) of the fiber-coupling process to realize the permanently coupled fiber-pigtailed QD micropillar. As described in Fig. 1(b), we apply the optical adhesive and UV curing for this purpose. Figure 3(a) shows the FC-EL emission spectra with the cavity modes $M_0$ and $M_1$ recorded for monitoring the coupling efficiency before, during and after the UV curing. The applied UV dose displayed in Fig. 3(b) was increased stepwise from $P_{\text{dose}} = 10$ mW/cm² to 160 mW/cm² during the 8 minutes curing period. During the 10 mW/cm² dose period, adjustment steps indicated with an arrow in Fig. 3(a) are visible in the spectra, which were applied to optimize the fiber-coupling efficiency using the FC-EL signal as feedback parameter. A comparison of the spectra before and after curing in Fig. 3(a) reveals a blue-shift of the cavity mode, which is observed with the increasing UV dose during the curing process, accumulating to a spectral shift of about -0.5 nm after curing is completed. This blue-shift, also observed in Ref. [19], can be explained by the strain applied to the micropillar by the fiber-ferrule due to shrinkage of the optical adhesive during the UV curing process. The observed mode-shift translates (via $\frac{\Delta\lambda}{\lambda} = \frac{\Delta n_{\text{eff}}}{n_{\text{eff}}}$ [29]) into a change of the effective refractive index of $\Delta n_{\text{eff}} = 0.0018$, which is even below the range of refractive index changes of $\Delta n = \pm 0.01$ typically expected in mechanically packaged waveguide structures [30]. Importantly, the magnitude of this shrinkage is not noticeably affecting the coupling efficiency of the device. After the UV curing step, the device was additionally encapsulated using epoxy glue. Comparing our curing process with the results reported by Haupt et al. in Ref. [20], we achieved a 14-times faster curing of our device combined with a negligible reduction in collected signal in the fiber during the coupling process. While the faster curing could be related to higher UV doses possible with our source based on collimated UV-LEDs, the noticeable decrease in signal intensity during curing in Ref. [20] in contrast to our results might result from the larger mode-field of their device reacting more sensitive to misalignments relative to the fiber-mode.

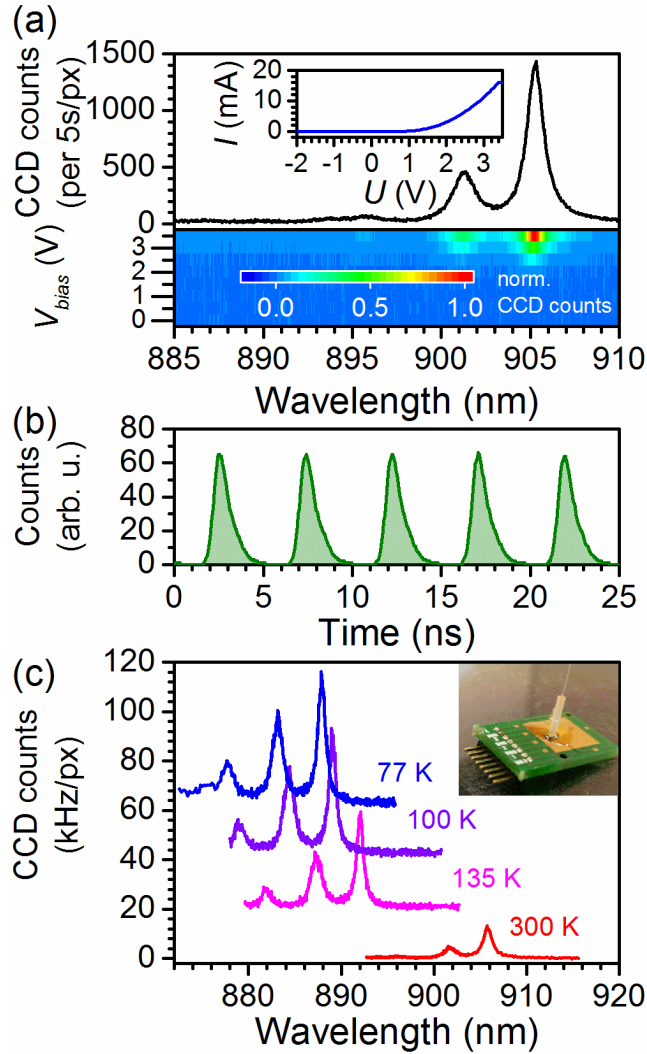

Figure 4. (a) FC-EL emission spectra of the fully fiber-coupled device after UV curing, with current-voltage characteristics as inset and FC-EL intensity as function of the applied dc bias (integration time for spectra: 5s). (b) Time-resolved FC-EL under pulsed current injection. (c) FC-EL spectra during cooldown of the fiber-coupled device. Inset: Exemplary picture of a fiber-coupled device.

Next, we investigate the electro-optical properties of the permanently fiber-pigtailed micropillar device, shown as inset of Fig. 4(c). Figure 4(a) presents the device's current-voltage characteristic under DC injection (~16 μA per pillar) as well as the integrated FC-EL emission intensity of the cavity mode $M_0$. The experimental data reveals a typical diode behavior of the fiber-pigtailed device, proving that the permanent coupling did not damage the wire bonds or the ring-shaped top-contact of the target device. The emission of the fiber-pigtailed micropillar after curing was monitored for several hours, and no change in emission intensity was observed, indicating a stable connection of fiber and sample.

Finally, to prove the applicability of our device for applications requiring electronic triggering, we measure the temporal response of the fiber-pigtailed micropillar under pulsed current injection using a silicon-based single photon counting module in combination with time-tagging electronics. The time-trace shown in Fig. 4(b) was recorded at a 206.2 MHz excitation repetition rate, a 6.4 V pulse amplitude and a 2.4 V dc offset. A periodic pulse train is observed with a decay time constant of 0.87 ns extracted from a mono-exponential fit to the falling pulse edge. To demonstrate the low-temperature suitability of our device, the fiber-pigtailed micropillar was transferred to a closed-cycle refrigerator with fiber-optical and electrical feedthroughs. Figure 4(c) shows EL-spectra under dc current injection of the device during the cooldown, starting from room-temperature down to a temperature of 77 K. The emission of the cavity modes is clearly visible in the entire temperature range, clearly demonstrating that low-temperature operation of our electrically driven SM fiber-coupled QD microcavity is possible. Note, however, that a direct judgment on the fiber-coupling efficiency is not possible from these data, due to the use of different bias-voltages and the relative spectral shift of cavity-modes and QD-ensemble. Reaching the temperature of 77 K thereby is an important achievement, allowing for simple and cost-effective cooling via liquid nitrogen or compact Stirling cryocoolers in future studies. While single-photon emission could not be demonstrated in the present fiber-coupled device, our results are very promising toward an all-fiber-coupled single-photon LED, as single QD effects can in principle be exploited in this temperature regime [31]. At temperatures below 75 K, we observed that the emission intensity became unstable and showed a noticeable degradation. While this behavior is not yet fully understood, we suspect the combination of UV adhesive and epoxy glue for encapsulation as a possible reason, requiring some more improvements/adjustments in the process.

In summary, we developed a technique enabling the direct and permanent coupling of electrically-pumped cavity-based QD light emitting devices to optical SM fibers. The coupling process, fully carried out at room temperature, involves a spatial scanning technique, where the fiber facet is positioned relative to a device with $d_{\text{pillar}} = 2.0$ µm using the fiber-coupled electroluminescence of the cavity emission as feedback parameter, ensuring optimum fiber alignment throughout the whole process. Subsequent gluing and UV curing enables a rigid and permanent coupling between micropillar and fiber core. Comparing our experimental results with FEM simulations indicates fiber coupling efficiencies of ~46% in our present device. Within a detailed characterization of the device after the fiber-coupling process, we demonstrate pulsed current injection as well as low-temperature operation, proving the suitability of our approach for quantum light generation in future experiments.

The next step is to demonstrate an efficient directly fiber-coupled single-photon source operated at cryogenic temperatures using the technique presented in this work. Considering that several publications proving single-photon emission [12,25,32] resulted from the same wafer material used in our present work, this seems to be feasible in the not too far future. The mode coupling efficiency, limited by the non-ideal mode-field matching between micropillar and fiber-core in the present study, can be substantially improved in the future by either employing fibers with smaller mode-field diameters or optimizing the layout of the QD-micropillar devices, both in combination with antireflection coatings on the fiber facet. A micropillar with $d_{\text{pillar}} = 6.0$ µm exhibits a simulated mode-radius of 2.54 µm, yielding a mode coupling efficiency of ~99% to the mode field of the given 780HP fiber. For a micropillar with $d_{\text{pillar}} = 2.0$ µm and the experimentally observed mode field radius in this work, an available speciality fiber (UHNA3/Nufern [33]) with a mode field radius below 1.3 µm could be used, yielding mode coupling efficiencies of ~91%. The technique for fiber-pigtailing cavity-enhanced QD light emitting diodes presented in this work is an important step in the quest for efficient and practical quantum light sources capable of high-speed electrical modulation [32] for applications in quantum information.

## ACKNOWLEDGMENTS

This work was financially supported by the German Federal Ministry of Education and Research (BMBF) via the project 'QuSecure' (Grant No. 13N14876) within the funding program Photonic Research Germany. Expert sample growth by M. Lermer as well as sample processing by M. Emmerling and A. Wolf is gratefully acknowledged. We further acknowledge Sven Burger and JCMwave GmbH for helpful discussions and support concerning the FEM simulations. The authors thank S. Reitzenstein for infrastructural support of this project.

## DATA AVAILABILITY

Data that support the findings of this study are available from the corresponding author upon reasonable request.

## REFERENCES

[1] J. L. O'Brien, A. Furusawa, and J. Vučković, Nat. Photonics 3, 687 (2009).
[2] I. Aharonovich, D. Englund, and M. Toth, Nat. Photonics 10, 631 (2016).
[3] A. Acín, I. Bloch, H. Buhrmann, T. Calarco, C. Eichler, J. Eisert, D. Esteve, N. Gisin, S. J. Glaser, F. J. Jelezko, S. Kuhr, M. Levenstein, M. F. Riedel, P. O. Schmidt, R. Thew, A. Wallraff, I. Walmsley, and F. K. Wilhelm, New J. Phys. 20, 080201 (2018).
[4] S. Rodt, S. Reitzenstein, and T. Heindel, J. Phys. Condens. Mat. 32, 153003 (2020).
[5] R. Trivedi, K. A. Fischer, J. Vučković, and K. Müller, Adv. Quantum Technol. 3, 1900007 (2020).
[6] H. Wang, Y.-M. He, T.-H. Chung, H. Hu, Y. Yu, S. Chen, X. Ding, M.-C. Chen, J. Qin, X. Yang, R.-Z. Liu, Z.-C. Duan, J.-P. Li, S. Gerhardt, K. Winkler, J. Jurkat, L.-J. Wang, N. Gergersen, Y.-H. Huo, Q. Dai, S. Yu, S. Höfling, C.-Y. Lu, and J.-W. Pan, Nat. Photonics 13, 770 (2019).

[7] S. Unsleber, Y.-M. He, S. Gerhardt, S. Maier, C.-Y. Lu, J.-W. Pan, N. Gregersen, M. Kamp, C. Schneider, and S. Höfling, Opt. Express. 24, 8539 (2016).

[8] X. Ding, Y. He, Z.-C. Duan, N. Gregersen, M.-C. Chen, S. Unsleber, S. Maier, C. Schneider, M. Kamp, S. Höfling, C.-Y. Lu and J.-W. Pan, Phys. Rev. Lett. 116, 20401 (2016).

[9] N. Somaschi, V. Giesz, L. De Santis, J. C. Loredo, M. P. Almeida, G. Hornecker, S. L. Portalupi, T. Grange, C. Antón, J. Demory, C. Gómez, I. Sagnes, N. D. Lanzillotti-Kumura, A. Lemaítre, A. Auffeves, A. G. White, P. Senellart, Nat. Photonics 10, 340 (2016).

[10] Z. Yuan, B. E. Kardynal, R. M. Stevenson, A. J. Shields, C. J. Lobo, K. Cooper, N. S. Beattie, D. A. Ritchie, and M. Pepper, Science 295, 102 (2002).

[11] C. L. Salter, R. M. Stevenson, I. Farrer, C. A. Nicoll, D. A. Ritchie, and A. J. Shields, Nature 465, 594 (2010).

[12] T. Heindel, C. A. Kessler, M. Rau, C. Schneider, M. Fürst, F. Hargart, W.-M. Schulz, M. Eichfelder, R. Roßbach, S. Nauerth, M. Lermer, H. Weier, M. Jetter, M. Kamp, S. Reitzenstein, S. Höfling, P. Michler, H. Weinfurter, and A. Forchel, New J. Phys. 14, 083001 (2012).

[13] M. Anderson, T. Müller, J. Huwer, J. Skiba-Szymanska, A. B. Krysa, R. M. Stevenson, J. Heffernan, D. A. Ritchie, and A. J. Shields, npj Quantum Inf. 6, 14 (2020).

[14] T. Kupko, M. v. Helversen, L. Rickert, J.-H. Schulze, A. Strittmatter, M. Gschrey, S. Rodt, S. Reitzenstein, and T. Heindel, npj Quantum Inf. 6, 29 (2020).

[15] H. Wang, Y. He, Y.-H. Li, Z.-E. Su, B. Li, H.-L. Huang, X. Ding, M.-C. Chen, C. Liu, J. Qin, J.-P. Li, Y.-M. He, C. Schneider, M. Kamp, C.-Z. Peng, S. Höfling, C.-Y. Lu, and J.-W. Pan, Nat. Photonics 11, 361 (2017).

[16] A. Schlehahn, S. Fischbach, R. Schmidt, A. Kaganskiy, A. Strittmatter, S. Rodt, T. Heindel, and S. Reitzenstein, Sci. Rep. 8, 1340 (2018).

[17] A. Musiał, K. Żołnac, N. Srocka, O. Kraverts, J. Große, J. Olszewski, K. Poturaj, G. Wójcik, P. Mergo, K. Dybka, M. Dyrkacz, M. Dłubek, K. Lauritsen, A. Bülter, P.-I. Schneider, L. Zschiedrich, S. Burger, S. Rodt, W. Urbańczyk, G. Sęk, and Stephan Reitzenstein, Adv Quantum Tech. 3, 2000018 (2020).

[18] X. Xu, I. Toft, R. T. Phillips, J. Mar, K. Hammura, and D. A. Williams, Appl. Phys. Lett. 90, 061103 (2007).

[19] D. Cadeddu, J. Teissier, F. R. Braakman, N. Gregersen, P. Stepanov, J.-M. Gérard, J. Claudon, R. J. Warburton, M. Poggio, and M. Munsch, Appl. Phys. Lett. 108, 11112 (2016).

[20] F. Haupt, S. S. R. Oemrawsingh, S. M. Thon, H. Kim, D. Kleckner, D. Ding, D. J. Suntrup III, P. M. Petroff, and D. Bouwmeester, Appl. Phys. Lett. 97, 131113 (2010).

[21] H. Snijders, J. A. Frey, J. Norman, V. P. Post, A. C. Gossard, J. E. Bowers, M. P. van Exter, W. Löffler, and D. Bouwmeester, Phys. Rev. Appl. 9, 031002 (2018).

[22] O. Ortiz, F. Pastier, A. Rodriguez, Priya, A. Lemaitre, C. Gomez-Carbonell, I. Sagnes, A. Harouri, P. Senellart, V. Giesz, M. Esmann, and N. D. Lanzillotti-Kimura et al., Appl. Phys. Lett. 117, 183102 (2020).

[23] L. Bremer, K. Weber, S. Fischbach, S. Thiele, M. Schmidt, A. Kaganskiy, S. Rodt, A. Herkommer, M. Sartison, S. L. Portalupi, P. Michler, H. Giessen, and S. Reitzenstein, APL Photonics 5, 106101 (2020).

[24] C. Böckler, S. Reitzenstein, C. Kistner, R. Debusmann, A. Löffler, T. Kida, S. Höfling, A. Forchel, L. Grenouillet, J. Claudon, and J. M. Gèrard, Appl. Phys. Lett. 92, 091107 (2008).

[25] T. Heindel, C. Schneider, M. Lermer, S. H. Kwon, T. Braun, S. Reitzenstein, S. Höfling, M. Kamp, and A. Forchel, Appl. Phys. Lett. 96, 011107 (2010).

[26] JCMsuite by JCMwave GmbH. The Simulation Suite for Nano Optics, 2021, https://jcmwave.com/ - The mode coupling efficiency is calculated as fraction of a TE dipole source's total power coupling to the two-fold degenerate fundamental mode of the modeled SM fiber. The TE dipole is situated in the center of the micropillar, resembling a QD.

[27] H. Rigneault, J. Broudic, B. Gayral, and J. M. Gérard, Opt. Lett. 26, 1595 (2001).

[28] A. Ghatak and K. Thyagarajan, Introduction to fiber optics (Cambridge University Press, Cambridge, 1998).

[29] J. Cai, Y. Ishikawa, and K. Wada, Opt. Exp. 21, 7162 (2013).

[30] M. Huang, Intern. J. of Solids and Struct. 40, 1615 (2003).

[31] A. Schlehahn, L. Krüger, M. Gschrey, J.-H. Schulte, A. Strittmatter, T. Heindel, S. Reitzenstein, Rev. Sci. Instr. 86, 013113 (2015)

[32] A. Schlehahn, A. Thoma, P. Munnelly, M. Kamp, S. Höfling, T. Heindel, C. Schneider, and Stephan Reitzenstein, APL Photonics 1, 011301 (2016)

[33] Nufern, Ultra-High NA Single Mode Fibers, 2020, https://www.nufern.com/pam/optical_fibers/spec/id/984/ - We point out that this fiber is specified for wavelengths of 960 nm or larger, making it not directly suitable for the emission wavelengths of the device used in this work. It emphasizes however, that technology is available for fibers with smaller mode fields, and we are of the opinion that it is sufficiently easy to either shift the device emission to larger wavelengths or use a similar fiber with slightly smaller wavelength requirements.